\documentstyle[aps,prd]{revtex} \input psfig.tex \voffset=.5truein
\begin{document} \title{Neutrino Absorption Tomography of the Earth's Interior
using Isotropic Ultra-high Energy Flux} \author{ Pankaj Jain$^a$, John P.
Ralston$^b$ and
George M. Frichter$^c$} \address{$^a$Physics Department, I.I.T., Kanpur, India
208016 \\ $^b$Department of Physics and Astronomy \\ University of Kansas,
Lawrence, KS 66045-2151, USA \\$^c$Department of Physics, Florida State
University, Talahassee, FL 32306-3016 USA, \\ } \date{\today} \maketitle

\begin{abstract} We study the feasibility of using an isotropic flux of cosmic
neutrinos in the energy range of 10 to 10000 TeV to study the interior
structure
of Earth. The angular distribution of events in a $\sim {\rm km}^3$-scale
neutrino telescope can be inverted to yield information on the Earth's mass
distribution that is independent of other methods.  The energy spectrum of
the neutrino primaries is also determined from
consistency with the angular distribution. 
It is possible to make a model
independent determination of the
density profile of Earth's interior, separate from the absolute
normalization of
the incident cosmic neutrinos. \end{abstract}

\section*{ }

The nature of the Earth's interior has traditionally been deduced by
indirect physical methods. An early, noteworthy result was Cavendish's 1793
deduction that the Earth must have a dense core, obtained by ``weighing the
Earth" gravitationally. 
Current measurements are based largely on seismic wave propagation, which
is rather indirect and has substantial intrinsic uncertainties
\cite{geotexts}.   Adding extra information fails to remove ambiguitites,
including studies ranging from the vibrational modes of the Earth as an
elastic body, to temperature constraints\cite{Jeanloz} , to the detailed
composition of the core \cite{Jeanloz}.   Controversies currently exist:
for example, seismic data has indicated that there may be an unsymmetrical
differentially rotating element in the core \cite{xsong}, with contenders
to explain this including a very large single crystal. Independent
measurements of the density profile would be of considerable value. Here we
discuss a novel way to take a rather direct `snapshot' of the nucleon
density in the Earth's interior, by considering tomography with ultra-high
energy neutrinos of cosmic origin. The principle of neutrino tomography is
essentially the same as X-ray tomography, except for substituting penetrating
neutrinos to serve in place of X-rays. By measuring neutrino absorption
along different paths through a solid body, one can deduce the nucleon
density in the
interior of the object. The results would be utterly independent of the
geophysical model, and directly measure the nucleon density.

The interaction strength of neutrinos with other fundamental particles
increases strongly with energy, and has been well measured at several high
energy
accelerators \cite{PDG98}.  The energy range of these direct measurements runs
from below $10^8$ eV up to almost $10^{14}$ eV, the latter being achieved
recently at the accelerator HERA in Hamburg.  For almost all of the
ultra-high energy (UHE)
region of incident energy above $10^{13}$ eV, the cross section has not been
measured directly. However, $\sigma$ can be calculated by exploiting relations
in the Standard Model between electron-initiated reactions which have been
measured, the neutrino initiated reactions desired, and evolution of active
quark and anti-quark pairs with energy \cite{Frichter95,Gandhi98}.  The
uncertainty due to theory in these calculations is small, leading to a
fundamental interaction which is sufficiently well known for the purposes of
tomography.

The flux of UHE neutrinos from cosmic sources cannot yet be considered
established, and several pilot experiments in the TeV (1 TeV = $10^{12}$ eV)
energy range are underway to measure it. The BAIKAL experiment operates in lake
water; the AMANDA project detects neutrinos interacting in the Antarctic ice
cap. A third scheme called RICE exists in the pilot stage, and uses a novel
radio detection strategy which is the most effective method above 100 TeV. The
optimal energy range for neutrino tomography is roughly 10-1000 TeV, a region
where these existing pilot projects have some overlap. However, current
detectors are small and would give a marginal or insufficient event rate for
Earth tomography.  With better resources, employing a few hundred optimally
tuned detectors, our calculations indicate that one should be able to say
something useful about the Earth's interior.  However our primary focus is
a future detector on a much larger scale, with a detection volume of order
1 ${\rm km}^3$ (KM3). With fluxes of the order of current astrophysical
estimates, a KM3 detector should be able to perform definitively.

Our approach differs from previous studies of neutrino absorption tomography.
The older studies concentrated on exploiting point neutrino sources assumed to
have a steady time dependence
\cite{Volkova74,DeRujula83,Wilson84,Askar'yan84,Borisov87,Nicolaidis91,Crawford95,Kuo95}.
Reconstruction of the density profile is
done by observing periodic occultation of the sources due to the Earth's
rotation. This method relies on the rate obtained from limited point
sources, and is also subject to errors if the energy dependence of the
primary spectrum is
poorly determined.  The geometry will not work for a detector located at
the South Pole.  Kuo {\it et al.} \cite{Kuo95} investigated Earth
tomography in the context of the DUMAND II array, concluding that a time
scale of `from years to decades' was needed to obtain sufficient data with
this
method. These results are important, but we offer a complementary and more
promising
scheme. We consider neutrinos coming from unresolved active galactic
nuclei, gamma ray bursts,
secondary emissions from cosmic rays whose directions are scrambled by
cosmic magnetic
fields, and other possible galactic or cosmological sources of ultra high
energy neutrinos.
Integrating over the Universe, this diffuse flux should be nearly
isotropic, with a sizable component in the optimal energy region. Such a
flux has several advantages. For example, one overcomes the serious problem
of binning events by arrival times to incorporate the effects of the
Earth's rotation.
There is far less ambiguity due to any possible time dependent fluctuations in
the flux of an energetic point source. Most attractively, the entire
Earth density profile can be obtained unambiguously, by a simple inversion of a
well-measured observable in the data, namely the angular distribution. The
overall normalization of the flux is not needed, as we arrange the
calculation so that it drops out of the determination of the density
profile.

The energy dependence of the primary neutrino flux is also determined by
the procedure.  This is unexpected and rather miraculous, but it occurs
because the interaction cross
section and detector efficiencies are energy dependent.  Given initial data on
the angular distribution, and supposing poorly determined initial data, or
guesses, on the
energy spectrum (always a problem in cosmic ray physics), our procedure
iterates the energy spectrum to obtain consistency with the angular
distribution and
density. Put another way, a faulty energy spectrum would be inconsistent,
and by
iteration the angular distribution and measured energy flux after
attenuation determine the energy spectrum. This is quite interesting and
may serve as a good method to measure the incident
energy spectrum. If one assumes that the density profile of the Earth is
already well
known, then the angular distribution strongly overdetermines the problem, and
one might even deduce the energy dependence of the cross section,
contributing a powerful check on fundamental physics.

To illustrate these remarks, consider the angular distribution of neutrinos
passing through the Earth (Fig. 1). With an isotropic primary flux, the angular
dependence comes from differing amounts of matter traversed en route to the
detector. The effect is expressed by an evolution equation for the flux,
$\Phi$, as a function of distance $z$ traversed:

\begin{equation} \frac{d {\rm ln}\Phi(E,z)}{dz}=-n(z)\sigma_{\rm eff}(E)\:.
\label{eqone} \end{equation}

Here $\sigma_{\rm eff}$ is a known `effective' cross section which incorporates
both charged current cross sections, neutral current cross sections, and
neutral current regeneration \cite{Berez86,Frich96} for neutrinos of energy
$E$. We
measure the polar angle $\theta$ with respect to the nadir (Fig. 1). We
assume spherical symmetry, so that the density $n=n(r)$ is a positive
definite function of distance
$r$ from the Earth's center. Let us assume momentarily that the measurement
is dominated by a
sufficiently narrow range of neutrino energies, so that the variation of
(\ref{eqone}) with energy can be neglected. By taking the logarithm of the
flux, the overall flux normalization is an additive constant that drops out
of the
angular distribution.

Solving (\ref{eqone}) for the surviving neutrino flux that can be measured at a
detector site located near the Earth's surface, one obtains,

\begin{equation} \Phi_{\rm surv}(E,\theta)=\Phi_\nu(E)e^{-\sigma_{\rm
eff}(E)R n(R) f(\theta)}\:, \label{eqtwo} \end{equation} where $\Phi_\nu$ is the
incident neutrino flux, $R$ is the Earth's radius  
and the function $f(\theta)$ is proportional to the  
integrated nucleon density along the chord $0<z<2R{\rm cos}(\theta)$. It is
convenient to measure $r$ in units of the Earth's radius $R$. Then,

\begin{equation} f(\theta)={1\over n(R)}\int_{{\rm sin}^2(\theta)}^1
\frac{n(r)d(r^2)}{\sqrt{r^2-{\rm sin}^2(\theta)}}\:. \label{eqthree}
\end{equation}
 Given data for the angular dependence over the region
$0<\theta<\frac{\pi}{2}$, this particular transform can be inverted; the
result is:

\begin{equation} n(r)=-{n(R)\over \pi}\int_{{\rm sin}^{-1}(r)}^{\frac{\pi}{2}}
\frac{df(\theta)}{d\theta} \frac{d\theta}{\sqrt{{\rm sin}^2(\theta)-r^2}}\:.
\label{eqfour} \end{equation}

This analytic result shows that the angular distribution is sufficient to
give the density profile.  The result is simpler than might be expected,
because the particular spherical geometry of the problem has been
exploited. Deviations from spherical symmetry are of interest, so that
relaxing our assumptions can be contemplated, but our goal here is to prove
the practicality of the simplest scheme when confronting realistic
difficulties.  The primary difficulty appears to be statistical
fluctuations from small expected data sets, but we will find that these
appear to be under control on the scale of KM3.

We now turn to the question of energy dependence that was sidestepped
above. The simple procedure above can be applied within a small energy bin.
One might then imagine requiring that in each angular bin, we also bin the
data in energy.  With limited
statistics and limited energy resolution of the detector, we find that such a
method is unlikely to be practical.  Yet integrating over energy does not
commute with the angular inversion, so a priori the energy integrated
angular distrbution does not appear to be adequate.  To get around these
problems, we created an alternative procedure in which the Earth's
density profile and the incident flux is iteratively improved.

We start by assuming a trial function for the density profile, which will yield
our initial guess for the attenuation factor $f(\theta)$. This can be used,
along with data on the energy dependence of observed flux integrated over
nadir angle,
to obtain our first guess for the incident energy spectrum. The attenuation
function
$f(\theta)$ can then be further improved by using the calculated value of the
incident flux.  Again this is compared with the angular distribution of the
observed flux integrated over energy. At no stage is it necessary to have the
joint distribution in energy and in angle.  This procedure is repeated till
convergence is obtained for both the incident flux and $f(\theta)$.  The same
procedure has then converged to the Earth's density profile. In order to obtain
convergence and a unique solution, it is necessary to fix one boundary
condition, which is taken to be the value of the density of the Earth near the
surface. Since the surface density is known with reasonable accuracy, the
boundary
condition should not introduce any bias.

In fact the result is overdetermined, because two moments of the density are
already known: These are the total mass of the Earth (just as Cavendish used),
and the Earth's moment of inertia.  But the entire procedure can be
carried out without making use of these moments, so in practice the density
is over-determined.  This is important, because one can expect only crude
measurement of the energy distribution from a realistic
cosmic ray detector.  

We used standard Monte Carlo methods to study the feasibility of the iterated
inversion technique. The simulations used the Preliminary Reference Earth Model
(PREM) \cite{Dzie81,geotexts} for the Earth's density profile. The range of
neutrino
energies was restricted to lie between $10 - 10^4$ TeV.  Our numerical
results show that the optimal lower limit in energy is between 10 and 50
TeV, since below this value the Earth is essentially
transparent. The optimal upper limit is between $10^3$ and $10^4$ TeV, beyond
which the number of events are expected to be too small to be of much use
for tomography.
We employed a generic form of the diffuse AGN neutrino flux,
$\Phi_\nu(E)=\Phi_oE^{-2}$ for $10\ {\rm TeV} < E < 10^4\ {\rm TeV}$. This
form is within the range of current theoretical predictions. For example,
in the energy range of interest, the AGN model of Stecker, Done, Salamon and
Sommers (SDSS) \cite{Stecker92} gives a flux $\sim E^{-1}$.  A model due to
Szabo and Protheroe (SP) \cite{Proth92}, while not now thought to be correctly
normalized, has the neutrino spectrum falling like $\sim E^{-2}$.  We take
an agnostic position on
the flux, and address uncertainties by simply renormalizing results at the
end. We simulated data for a generic UHE neutrino telescope, for the
purposes of study defined in two ways: in one extreme for simplicity, we
took detector response independent of neutrino energy and angle of
incidence. The other extreme is the case of detector with the energy and
angular response calculated for a radio array \cite {Frich96}.  The radio
method has an response strongly increasing with energy, making the flat
response to the higher-energy part of the spectrum a more conservative
method. 
Meanwhile we do not have sufficient information on 
the angular
response of optical detection. Since we believe that a realistic detection
scheme would combine the strengths of both optical and radio detection,
the two cases should give a reasonable range of results without excluding
either or 
getting bogged down in details: in fact, the results were so similar 
that we simply report the simpler (isotropic and flat) response. 
Of course, detector response for a particular experimental situation, as
well as realistic energy resolution and pointing accuracy, can always be
incorporated.

Our simulation generated $N$ events distributed in nadir angle,
$0^o<\theta<90^o$, and energy, 10 TeV $<E_\nu< 10^4 $ TeV, according to
Eqn.  (\ref{eqtwo}).  We normalized our calculation so the total number of
events observed per antenna between 100 TeV and $10^4$ TeV is about 100 per
year. This is about one fourth of the event rate calculated  in
\cite{Frich96}, which takes into account subsequent changes in estimates of
the incident flux.  While we thus normalize our rates to radio detection,
any combination of methods can be rescaled in an obvious way.

The Monte Carlo data were divided into 20 energy $E$ bins, with widths
increasing like $E^2$. The data was also divided into 10 angular bins
chosen to be equally spaced in radius from the center of the Earth. These
bins were chosen to obtain the density at roughly uniform intervals. We
made no attempt to discover an optimal binning procedure. However the solid
angle subtended by the central bin, namely the one containing the center of
Earth, is very small, and hence it puts severe requirements on the
total number of events needed to say something useful about the density in
this region.
This also happens to be one of the most interesting regions for geophysics.
Depending on the total number of events and optimization scheme, one may
wish to adjust this bin to get a better measurement.

We found that for a wide range of trial density profiles convergence of the
iterative procedure was obtained within about 5 iterations.  We plot in
Fig. 2 an example of successive approximations to the attenuation function
$f(\theta)$, with the solid line showing the final result. Fig. 3
shows the successive approximations for the incident flux in energy.  The
plots show rapid convergence occurs with the scheme chosen. The figures
show explicitly that a considerable error in
the incident energy spectrum can be tolerated, with the final energy
spectrum converging
to the actual spectrum.

The final extracted density profile, along with the PREM density profile,
is shown in
Fig. 4.  The step between the core and lower mantle is very well resolved,
while the inner core is not.  Since the density converged to proper value,
then all of its moments also converged, showing that the total mass and
moment of inertia were consistent, or: Neutrinos can ``weigh the Earth".
In practice the known mass density moments provide an excellent handle on
the overall consistency of the final result.  Statistical error bars on the
derived density are acceptable (Fig. 4), although one would probably want
to optimize the central core region further. The results were obtained by
assuming that a detector with 1000 antennas is deployed for two years.  The
number is ambitious but within the scope of planning for future arrays.
Because the errors are statistical, the same result would be obtained
for an incident flux normalized 5 times lower in 10 years running.  Put yet
another way, even a modest array of 200 antennas might say something useful
in the time scale of 10 years. In Fig. 5 we also show the final result
assuming the optimistic flux estimates, but only 100 antennas for a running
time of 2 years. In this case we divide the earth's radius into only five
bins instead of ten, in order to get a reasonable number of events in the
central bin. The step between the core and lower mantle remains well
resolved.

We hasten to add that there can be many uncertainties in realistic
experimental design, which only further study can address.  During a period
of a few years to a decade, a $KM3$ neutrino telescope will be fulfilling a
primary mission as a fundamentally new kind of instrument for observing the
cosmos. We believe that with the same kind of detector, neutrino tomography
could also provide important and independent information about the Earth's
interior.

{\large\bf Acknowledgements:} We thank Geoff Abers, John Doveton, Doug McKay
and R. P. Singh for useful comments.
Supported
by DOE grant number DE-FGO2-98ER41079, the KU General Research Fund, NSF-K*STAR
Program under the Kansas Institute for Theoretical and Computational
Science and DAE grant number DAE/PHY/96152.

\newpage
\begin{figure} [t,b] \hbox{\hspace{4em}
\hbox{\psfig{file=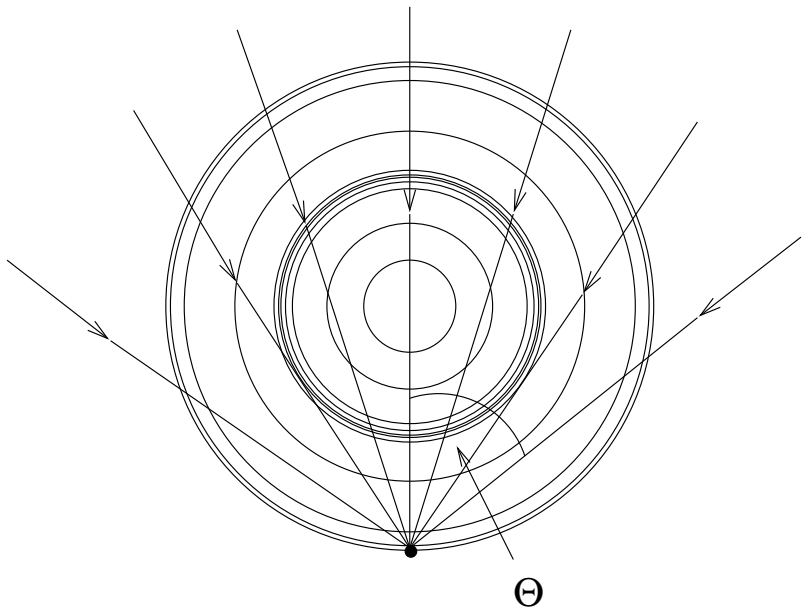,height=8cm}}}
\bigskip
\caption
{Earth density contours, and paths traced by netrinos at different angles
of incidence to the detector.
}
 \label{fig1}
\end{figure}

\newpage
\begin{figure} [t,b] \hbox{\hspace{4em}
\hbox{\psfig{file=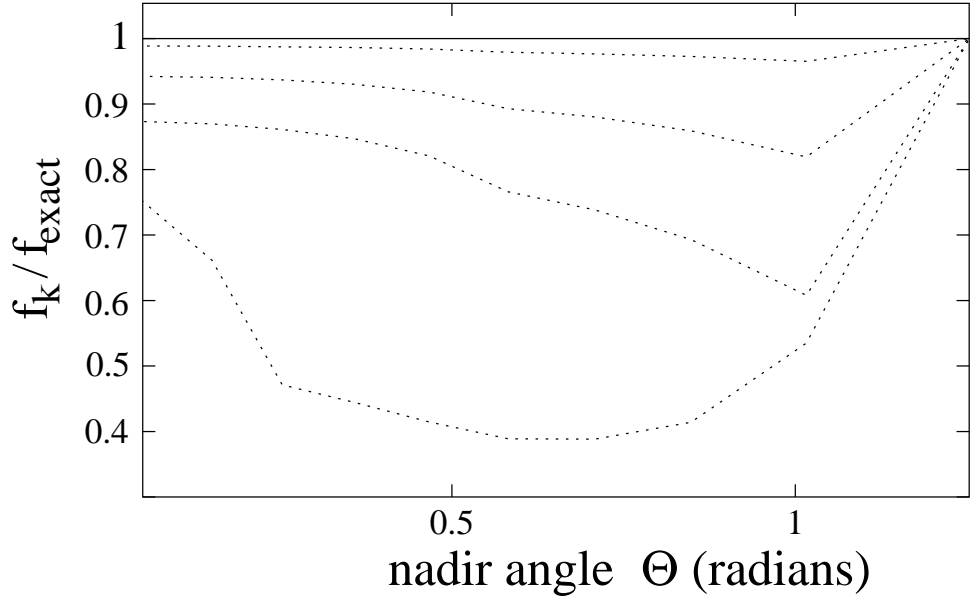,height=8cm}}}
\bigskip
\caption{Successive approximations to the attenuation function $f(\theta)$
calculated iteratively, where $\theta$ is the nadir angle. The solid line
shows the final result} \label{fig2}
\end{figure}

\begin{figure} [t,b] \hbox{\hspace{4em}
\hbox{\psfig{file=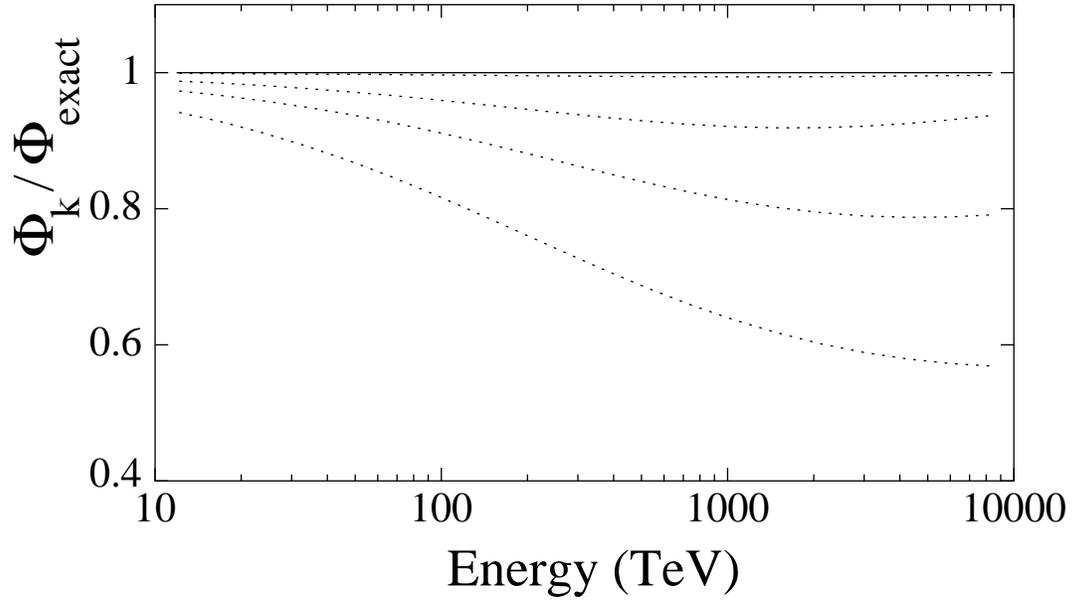,height=8cm}}}
\bigskip
\caption{Successive approximations to the extracted value of the incident
flux $\phi(E)$, calculated
iteratively. The solid line shows the final result.} \label{fig3} \end{figure}

\begin{figure} [t,b] \hbox{\hspace{4em}
\hbox{\psfig{file=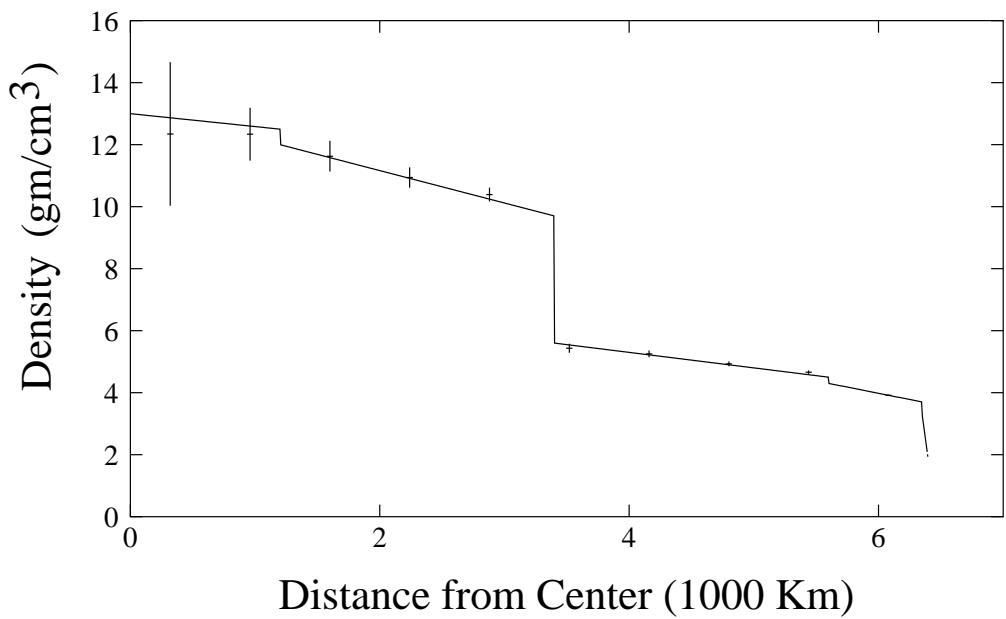,height=8cm}}}
\bigskip
\caption{The extracted density using ten radial bins along with the
statistical errors. The PREM model for the Earth's density, solid curve,
was used to generate data. The step between the core and lower mantle is
very well resolved, while the difference between the inner and outer cores
is not.} \label{fig4} \end{figure}

\begin{figure} [t,b] \hbox{\hspace{4em}
\hbox{\psfig{file=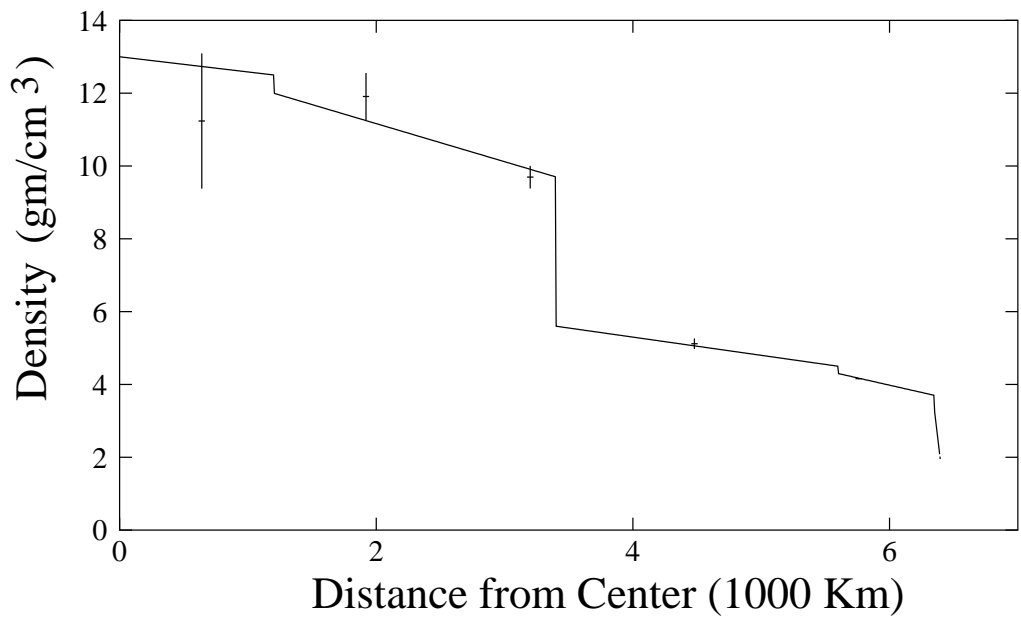,height=8cm}}}
\bigskip
\caption{The extracted density using five radial bins along with the
statistical errors, assuming one tenth the event rate compared to Fig. 4.
The step between the core and lower mantle remains well resolved. }
\label{fig5} \end{figure}
\end{document}